\documentstyle[11pt,bezier]{article}
\renewcommand{\baselinestretch}{1.2}
\def\ni{\noindent}
\begin{document}

\newpage
\null


\begin{center}
\section*{HIGHLIGHT OF THE MONTH}
\end{center}
\vspace{1cm}
\rule{16.5cm}{1mm}
\vspace{1cm}
\begin{center}
{\LARGE {\bf 3-Body Scattering (3BS) theory of 
on-site correlation in narrow band materials}}\\

\vspace*{4mm}

{\large{ Franca Manghi and Massimo Rontani}}\\
{\it Istituto Nazionale per la Fisica della Materia 
- Dipartimento di Fisica, Universit\`a di Modena 
Via Campi 213/a I-4110 Modena Italy}\\
\end{center}
\vspace{1cm}
\rule{16.5cm}{1mm}
\vskip 0.5cm

\begin{abstract}
We present the results of a recently developed approach where the interplay
between the itinerant and localized character of electrons in narrow band
materials is described by adding on-site correlation effects to a
first principle band calculation: the single particle band states are
treated as mean field solutions of a multi-orbital Hubbard Hamiltonian and
the many-body term associated to localized e-e interaction is 
described in a configuration-interaction scheme.
The method allows to calculate hole and electron 
spectral functions which can be directly 
compared with spectroscopical results. 
\end{abstract}

\subsection*{Introduction}

The Hubbard model, dominated by the competition between inter-site hopping
and on-site electron-electron repulsion, is believed to describe the
physics of narrow band materials such as transition metals, transition
metal oxides, cuprates, etc.~\cite{Dagotto}. In these systems the itinerant
character of valence electrons - clearly shown by the 
${\bf k}$-dispersion observed
in photoemission - coexists with strong local correlations responsible of
other spectroscopical features - satellites, band-narrowing, and opening,
in some cases, of a Mott-Hubbard gap. 

In spite of the enormous amount of theoretical and experimental work which
has been done on cuprates since the discovery of high $T_c$
superconductors, an unified theoretical description of the whole valence
spectrum, from the high binding energy region characterized by satellites,
up to the valence band top, including both unperturbed single particle like
and strongly correlated Cu derived structures, is still missing; this is
due to the difficulty to combine an accurate treatment of many body terms
with a realistic description of the band structure. 

Most of the work on this subject has been based on a drastic simplification
of either the band structure or the e-e interaction; if the complex
structure of these systems is reduced to a model description involving only
CuO planes, the e-e interaction can be treated accurately, for instance  by
exact diagonalization techniques of finite (and small) CuO clusters
\cite{Dagotto}, by refined mean-field solutions of a 
two dimensional {\it t-J}
Hamiltonian \cite{Kotliar86} etc. . However, in this search for the
simplest model containing the relevant physics of superconductors one may
miss some important effects, related for instance to the coupling between
adjacent CuO planes \cite{Biagini95}, and the possibility of a quantitative
comparison with spectroscopical results. 

Photoemission data of highly correlated materials have been also interpreted
using theoretical approaches based on impurity and cluster
configuration-interaction models which assume a strong wave function
localization and adopt a rather simplified description of the band
structure, with a considerable number of adjusting parameters. They have
been widely used to describe the main structures and satellite peaks
observed in the angle-integrated photoemission spectra of CuO
\cite{Sawatzky} and of cuprate superconductors \cite{Shen}. Other
approaches have been proposed based on the  density functional
approximation (self-interaction corrected \cite{Svane} and  LDA+U
\cite{Anisimov} functionals) which fully include the itinerant character of
electron states but describe the electron-electron interaction as a
mean-field effective single-particle potential. 

A theoretical approach is then needed which includes both the hybridization
between Cu and the ligands (or between 
{\it sp} and {\it d} states in the case of
transition metals) accounted for by first principle band theory, and a
treatment of e-e interaction which must be non-perturbative - to deal with
systems which are in the high correlation regime - and beyond mean field -
to include finite life-time excitations. 

The 3BS method \cite{Igarashi,Calandra,Manghi97,Fulde} can be seen as an
extension to the solid state of the configuration-interaction scheme used
for finite systems: the Hubbard Hamiltonian is projected on a set of states
obtained by adding a finite number of e-h pairs to the ground state  of the
single-particle Hamiltonian and this expansion is truncated to include one
e-h pair. The effect of electron correlation on one electron removal
energies from a partially filled band is then described as hole-hole and
hole-electron interaction. The 3BS theory corresponds to the solution of a
3-body scattering problem involving two holes and one electron and has been
originally formulated by Igarashi \cite{Igarashi}. Self-energy corrections,
spectral functions and quasi-particle band structure can be calculated for
systems in different correlation regimes, getting a complete picture of the
whole valence spectrum, including both long-lived coherent quasiparticle
structures and incoherent short-lived ones. 

\subsection*{Hamiltonians} 

Since we want to augment band theory with the inclusion of on-site 
correlation it is essential to define the relationship between 
band and Hubbard Hamiltonian in order to avoid double counting of 
e-e interaction. 
The exact many body Hamiltonian in second quantization is 
\begin{eqnarray}
\label{acca1}
\hat{H} &=&
\sum_{i\alpha\sigma} \epsilon_{i\alpha} \hat{n}_{i\alpha\sigma}+
\sum_{\alpha \beta \sigma} {\sum_{ij}}^{\prime} 
t_{i \alpha, j \beta} \hat{c}_{i \alpha \sigma}^{\dagger} 
\hat{c}_{j \beta \sigma} \nonumber \\
&+&
\frac{1}{2}
\sum_{ \alpha \beta} \left [ \sum_{i} (U_{\alpha \beta} - J_{\alpha \beta})
\sum_{\sigma} \hat{n}_{i\alpha\sigma} \hat{n}_{i\beta\sigma}+
\sum_{i} U_{\alpha \beta} 
\sum_{\sigma} \hat{n}_{i\alpha\sigma} \hat{n}_{i\beta -\sigma}\right ]
\nonumber \\
&+&
...... \mbox{(multi-center terms)},
\end{eqnarray}
with $\hat{n}_{i\alpha\sigma}=\hat{c}_{i\alpha\sigma}^{\dagger}
\hat{c}_{i\alpha\sigma}$ and 
$\hat{c}_{i\alpha\sigma}, \hat{c}_{i\alpha\sigma}^{\dagger}$
destruction and creation operators.

Here $\epsilon_{i\alpha}$ and $t_{i\alpha , j\beta}$ are the 
intra- 
and inter-atomic matrix elements of the one-particle Hamiltonian and 
$U_{\alpha \beta}$, $J_{\alpha \beta}$ are 
on-site Coulomb and exchange terms:
\[
U_{\alpha \beta} = 
V_{i\alpha\sigma,i\beta\sigma,i\beta\sigma,i\alpha\sigma}=
V_{i\alpha\sigma,i\beta -\sigma,i\beta -\sigma,i\alpha\sigma},
\]
\[
J_{\alpha \beta} = 
V_{i\alpha\sigma,i\beta\sigma,i\alpha\sigma,i\beta\sigma},
\]
with
\[
V_{i\alpha\sigma,j\beta\sigma^{\prime},l\gamma\sigma^{\prime},m\delta
\sigma}=\sum_{s s'}\int\!\!\phi_{i\alpha\sigma}^{*}({\bf r},s)
\phi_{j\beta\sigma^{\prime}}^{*}({\bf r}^{\prime},s^{\prime})
\frac{e^2}{|{\bf r}-{\bf r}^{\prime}|}
\phi_{l\gamma\sigma^{\prime}}({\bf r}^{\prime},s^{\prime})
\phi_{m\delta\sigma}({\bf r},s){\rm d}\,{\bf r}{\rm d}\,{\bf r}^{\prime}.
\]

Different approximations to the exact Hamiltonian (\ref{acca1}) can be
obtained using a mean field approach which amounts to neglect
fluctuations in the electron occupation 
\begin{eqnarray*}
\hat{n}_{i\alpha\sigma} \hat{n}_{i\beta \sigma'} &=& 
\hat{n}_{i\alpha\sigma} \left<\hat{n}_{i\beta \sigma'}\right> +
\hat{n}_{i\beta\sigma'} \left<\hat{n}_{i\alpha \sigma}\right> -
\left<\hat{n}_{i\alpha\sigma}\right> 
\left<\hat{n}_{i\beta \sigma'}\right> \nonumber \\
&+& 
\left(\hat{n}_{i\alpha\sigma}-\left<\hat{n}_{i\alpha \sigma}
\right>\right)\left(\hat{n}_{i\beta\sigma'}-\left<
\hat{n}_{i\beta \sigma'}\right>\right) \nonumber \\
&\simeq&
\hat{n}_{i\alpha\sigma}\left<\hat{n}_{i\beta \sigma'}\right> +
\hat{n}_{i\beta\sigma'}\left<\hat{n}_{i\alpha \sigma}\right> -
\left<\hat{n}_{i\alpha\sigma}\right>\left<\hat{n}_{i\beta \sigma'}\right>.
\end{eqnarray*}
The mean field 
approximation can be applied to all the many body terms of (\ref{acca1})
transforming it into a single particle Hamiltonian
\begin{equation}
\label{accamf}
\hat{H}^{MF}=
\sum_{i\alpha\sigma} 
\epsilon^{MF}_{i \alpha \sigma} \hat{n}_{i\alpha\sigma}+
\sum_{\alpha \beta \sigma} {\sum_{ij}}^{\prime} 
t_{i \alpha, j \beta} \hat{c}_{i \alpha \sigma}^{\dagger} 
\hat{c}_{j \beta \sigma},
\end{equation}
or selectively to 
the multi-center integrals, keeping the full many body character in 
the one-center terms; in this way one gets the generalized Hubbard model
\begin{eqnarray}
\hat{H}^{H} &=&
\sum_{i\alpha\sigma} \epsilon^H_{i\alpha\sigma} \hat{n}_{i\alpha\sigma}+
\sum_{\alpha \beta \sigma} {\sum_{ij}}^{\prime} 
t_{i \alpha, j \beta} \hat{c}_{i \alpha \sigma}^{\dagger} 
\hat{c}_{j \beta \sigma} \nonumber \\
&+&
\frac{1}{2}
\sum_{ \alpha \beta}\left [ \sum_{i} 
\left(U_{\alpha \beta} - J_{\alpha \beta}\right)
\sum_{\sigma} \hat{n}_{i\alpha\sigma} \hat{n}_{i\beta\sigma}+
\sum_{i} U_{\alpha \beta} 
\sum_{\sigma} \hat{n}_{i\alpha\sigma} \hat{n}_{i\beta -\sigma} \right].
\end{eqnarray}
Since $\hat{H}^{MF}$ and $\hat{H}^{H}$ 
differ only for the treatment of the on-site 
correlation - included in $\hat{H}^{MF}$ as a mean field and treated as a many 
body term in $\hat{H}^H$ - it is easy to show that 
\begin{equation}
\label{emf}
\epsilon^{MF}_{i \alpha \sigma} =
\epsilon^{H}_{i\alpha\sigma} +
\sum_{  \beta} \left[\left(U_{\alpha \beta} - J_{\alpha \beta}\right)
\left<\hat{n}_{i\beta\sigma}\right> +
U_{\alpha\beta}\left<\hat{n}_{i\beta -\sigma}\right>\right]. 
\end{equation}

Using a  Bloch basis set the two approximate Hamiltonians become
\begin{eqnarray}
\label{hhk}
\hat{H}^H  &=&  
\sum_{{\bf k} n \sigma}\epsilon^H_{{\bf k}n\sigma}
\hat{a}^{n^{\dagger}}_{{\bf k}\sigma}\hat{a}^n_{{\bf k}\sigma} 
+ \sum_{\alpha\beta}
\sum_{{\bf k}{\bf k}'{\bf p}}
\sum_{n n'} \sum_{m m'}
\sum_{\sigma}
\frac{1}{2N} \nonumber \\
& &\times \left[U_{\alpha\beta}
C_{\alpha \sigma}^n({\bf k})^{*}
C_{\alpha \sigma}^{n'}({\bf k}+{\bf p})
C_{\beta -\sigma}^m({\bf k}')^{*}
C_{\beta -\sigma}^{m'}({\bf k}'-{\bf p})
\hat{a}^{n\dagger}_{{\bf k} \sigma}
\hat{a}^{ n'}_{{\bf k}+{\bf p} \sigma}
\hat{a}^{m\dagger}_{{\bf k}'-\sigma}
\hat{a}^{m'}_{{\bf k}'-{\bf p} -\sigma}\right. \nonumber \\
&+&
\left.\left(U_{\alpha\beta}-J_{\alpha\beta}\right)
C_{\alpha \sigma}^n({\bf k})^{*}
C_{\alpha \sigma}^{n'}({\bf k}+{\bf p})
C_{\beta \sigma}^m({\bf k}')^{*}
C_{\beta \sigma}^{m'}({\bf k}'-{\bf p})
\hat{a}^{n\dagger}_{{\bf k} \sigma}
\hat{a}^{ n'}_{{\bf k}+{\bf p} \sigma}
\hat{a}^{m\dagger}_{{\bf k}'\sigma}
\hat{a}^{m'}_{{\bf k}'-{\bf p} \sigma} \right],
\end{eqnarray}
\begin{equation}
\label{hbk}
\hat{H}^{MF}  =  
\sum_{{\bf k} n \sigma}\epsilon^{MF}_{{\bf k}n\sigma}
\hat{a}^{n^{\dagger}}_{{\bf k}\sigma}\hat{a}^n_{{\bf k}\sigma};
\end{equation}
here $\hat{a}^n_{{\bf k}\sigma}$,$\hat{a}^{n^{\dagger}}_{{\bf k}\sigma}$
are destruction/creation operators of electrons with wave vector ${\bf k}$
, spin $\sigma$, band index $n$ and $C^n_{\alpha}({\bf k}\sigma)$ are
the expansion coefficients of Bloch states in terms of localized orbitals. 

The relationship between single particle eigenvalues 
$\epsilon^{MF}_{{\bf k}n\sigma}$ and $ \epsilon^{H}_{{\bf k}n\sigma} $
appearing in the two Hamiltonians is now 
\begin{equation}
\label{eek}
\epsilon^{MF}_{{\bf k}n\sigma} = \epsilon^{H}_{{\bf k}n\sigma} +
Q^n_{{\bf k} \sigma},
\end{equation}
\begin{equation}
\label{qmf}
Q^n_{{\bf k} \sigma}= \sum_{\alpha \beta} 
|C_{\alpha \sigma}^n({\bf k})|^2 \left[ U_{\alpha \beta}\frac{1}{N} 
\sum^{occ}_{{\bf k'} n'} 
|C_{\beta -\sigma}^{n'}({\bf k'})|^2 + \left(U_{\alpha \beta}
-J_{\alpha \beta}\right)\frac{1}{N} 
\sum^{occ}_{{\bf k'} n'}|C_{\beta \sigma}^{n'}({\bf k}')|^2 \right], 
\end{equation}
which is the analogue of eq. (\ref{emf}) for Bloch 
states. 
Any band structure calculation corresponds to the solution of 
some $\hat{H}^{MF}$ describing the interacting system as an effective 
single particle problem and equations (\ref{eek},\ref{qmf})
contain the correct recipe to include 
Hubbard correlation starting from band structure eigenvalues, avoiding e-e
interaction double counting. 

\subsection*{Hole spectral function,  self-energy and the Faddeev method}
\label{sec-selfe}

We are interested in the hole spectral function 
\begin{equation}
\label{specf}
D^{-}_{{\bf k}  \sigma}(\omega) =  \frac{1}{\pi} 
\sum_{n} {\rm Im}\, {\cal G}^-\!\left({\bf k} n \sigma,\omega\right),
\end{equation}
which describes the removal of one electron 
of wave-vector ${\bf k}$, band index $n$ and spin $\sigma$ and is related 
to the hole-propagator
\begin{eqnarray}
\label{GH}
{\cal G}^-\!\left({\bf k} n \sigma,\omega\right)&=&
-\left<\Psi_0\right|\hat{a}^{n^{\dagger}}_{{\bf k}\sigma}
\hat{G}\left(z\right)
\hat{a}^n_{{\bf k}\sigma} \!\left|\Psi_0\right>  , \qquad
\qquad z=-\omega+E_0\left(N_e\right)+{\rm i}\delta; \nonumber\\
\end{eqnarray}
$E_0\left(N_e\right)$ and $\left|\Psi_0\right>$ define the ground 
state of the $N_e$ particle system and 
\begin{equation}
\hat{G}\left(z\right)=\frac{1}{z-\hat{H}^H}
\label{resolvent}
\end{equation}
is the resolvent operator. 
By projecting the Hamiltonian (\ref{hhk}) over a complete set 
appropriate for the $N_e-1$ particle system one gets an 
expression for $\hat{H}^H$  appropriate to describe one electron removal.
The key approximation is to choose  a subset of all the excited states 
of the {\it non-interacting} system and assume it to be complete.
Any N particle non-interacting state can be obtained by repeated 
applications of creation/destruction operators to the ground state Slater 
determinant $\left|\Phi_0\right>$ i.e.~by adding e-h pairs to it; according to
3BS the {\it interacting} state with one removed electron of momentum ${\bf k}$ 
and spin $\sigma$  is expanded in terms of the basis set including 
1-hole and 3-particle configurations
\begin{equation}
\left|s\right> \equiv \hat{a}_{{\bf k} n \sigma} \left|\Phi_0\right>,
\ \ \ \
\left|t\right> \equiv
\hat{a}^{\dagger}_{{\bf q}_3 n_3 \sigma_3} \hat{a}_{{\bf q}_2 n_2 \sigma_2}
\hat{a}_{{\bf q}_1 n_1 \sigma_1}\left|\Phi_0\right>, 
\label{states}
\end{equation}
with 
\[
{\bf q}_1+{\bf q}_2-{\bf q}_3= {\bf k}, \ \  \qquad
\qquad \sigma_1+\sigma_2-\sigma_3 = \sigma. 
\]
The effective Hamiltonian for the N-1 particle system is then
\begin{equation}
\hat{H}^H_{N_e-1} \simeq \hat{H}_1 + \hat{H}_3 +\hat{V},
\end{equation}
where $\hat{H}_1$ is associated to one-hole configurations
\[
\hat{H}_1 = \left<s\right|\hat{H}^H\left|s\right> 
\left|s\right>\left<s\right|,
\]
$\hat{H}_3$
describes the contribution of 3-particle configurations 
\[
\hat{H}_3= 
\sum_{t t'} 
\left<t\right|\hat{H}^H\left|t'\right> 
\left|t\right>\left<t'\right|,
\]
and $\hat{V}$ is the coupling between 1- and 3-particle states
\[
\hat{V} = \sum_{t} 
\left<s\right|\hat{H}^H\left|t\right> 
\left|s\right>\left<t\right| + h.c.\ \ .
\]
We can now calculate the resolvent (\ref{resolvent}). We define the
3-particle resolvent, that is the resolvent associated to the 3-particle
interaction 
\[
\hat{F}_3(z) =  \frac{1}{z-  \hat{H}_3},
\]
and the Dyson equation which relates $\hat{G}(z)$ to it
\begin{equation}
\hat{G}(z)=\hat{F}_3(z)+\hat{F}_3(z) [\hat{H}_1+\hat{V}]\hat{G}(z).
\label{dyson}
\end{equation}
The Faddeev scattering theory allows to determine $\hat{F}_3(z)$ by 
separating the 3-body Hamiltonian in diagonal and non-diagonal parts 
\[
\hat{H}_3^D=
\sum_{t} 
\left<t\right|\hat{H}^H\left|t\right> 
\left|t\right>\left<t\right|,
\]
\[
\hat{H}_3^{ND}=
{\sum_{t t'}}^{\prime} 
\left<t\right|\hat{H}^H\left|t'\right> 
\left|t\right>\left<t'\right|,
\]
defining the {\it diagonal} 3-body resolvent
\[
\hat{F}_3^D\left(z\right)=\frac{1}{z-\hat{H}_3^D},
\]
and the scattering operator 
\[
\hat{S}=\hat{H}_3^{ND} + \hat{H}_3^{ND} \hat{F}_3^D \hat{S}.
\]
The {\it full} 3-body resolvent can be written in terms of the diagonal one and 
of the scattering operator as 
\begin{equation}
\label{f3}
\hat{F}_3 = 
\hat{F}_3^D + 
\hat{F}_3^D   \hat{S}  \hat{F}_3^D.
\end{equation}
As shown in references \cite{Calandra,Manghi97} the non-diagonal
3-body interaction is the sum of two potentials 
\[
\hat{H}_3^{ND} = \hat{V}_{h-h} + \hat{V}_{h-e},
\]
which describe h-h and h-e multiple scattering.
%
%
\begin{figure}[thb]
\begin{center}
 \setlength{\unitlength}{0.1cm}
 \begin{picture}(110,75)
 \put(25,5){\line(0,1){5}}
 \thicklines
 \put(5,40){\line(0,1){2}} 
 \put(25,10){\line(0,1){35}}
 \put(25,47){\line(0,-1){2}}
 \put(45,47){\line(0,-1){2}}	
 \put(5,42){\line(1,0){20}}
 \put(25,47){\line(1,0){20}}
 \thinlines
 \put(5,40){\line(1,0){40}}
 \put(25,45){\line(0,1){10}}
 \thicklines
 \put(15,50){\vector(0,1){5}}
 \put(35,55){\vector(0,-1){5}}
 \put(15,30){\circle{5}}
 \bezier{250}(25,10)(5,10)(5,40)
 \bezier{250}(25,15)(45,15)(45,45)
 \multiput(42,20)(4,0){6}{\line(1,0){3}}
 \put(52,22){ $\hat{V}$}
 \put(39,20){\vector(-1,0){0}}
 \put(68,20){\vector(1,0){0}}
 \put(85,5){\line(0,1){5}}
 \thicklines
 \put(85,10){\line(0,1){37}}
 \put(65,42){\line(1,0){20}}
 \put(65,42){\line(0,-1){2}}
 \put(85,47){\line(1,0){20}}
 \put(105,47){\line(0,-1){2}}
 \thinlines
 \put(85,40){\line(1,0){20}}
 \put(65,40){\line(1,0){20}}
 \put(72,23){$\hat{V}_{\rm h-h}$}
 \put(89,33){$\hat{V}_{\rm h-e}$}
 \put(85,45){\line(0,1){10}}
 \thicklines
 \put(75,50){\vector(0,1){5}}
 \put(95,55){\vector(0,-1){5}}
 \put(79,35){\vector(2,1){12}}
 \put(79,35){\vector(-2,-1){0}}
 \put(79,31){\vector(3,-2){9}}
 \put(79,31){\vector(-3,2){0}}
 \put(75,33){\circle{5}}
 \put(92,23){\circle{5}}
 \put(95,43.5){\circle*{5}}
 \bezier{250}(85,10)(65,10)(65,40)
 \bezier{250}(85,15)(105,15)(105,45)
 \end{picture}
 \end{center}
\caption{
Pictorial representation of non-diagonal terms in the effective hole
Hamiltonian: $\hat{V}$ couples one- and three-particle configurations while
$\hat{V}_{h-h}$ and $\hat{V}_{h-e}$ describe scattering between
three-particle states, namely hole-hole and hole-electron scattering
respectively. 
}\label{fig_1}
\end{figure}
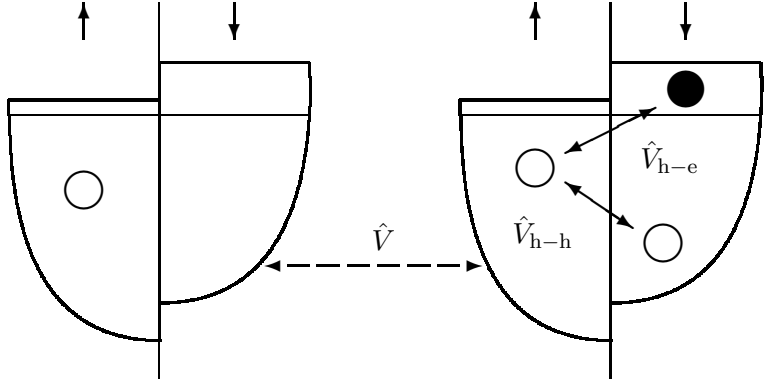
We define partial scattering operators $\hat{S}_{h-h}$,$\hat{S}_{h-e}$
such that $\hat{S} = \hat{S}_{h-h}+\hat{S}_{h-e}$, i.e.
\begin{eqnarray*}
\hat{S}_{h-h} &=& \hat{V}_{h-h} + \hat{V}_{h-h} \hat{F}_3^D \hat{S}, \\ 
\hat{S}_{h-e} &=& \hat{V}_{h-e} + \hat{V}_{h-e} \hat{F}_3^D \hat{S},
\end{eqnarray*}
which are related to the scattering T-matrices 
\label{tmat}
\begin{equation}
\hat{T}_{h-h} = \hat{V}_{h-h} + \hat{V}_{h-h} \hat{F}_3^D \hat{T}_{h-h}, \\
\end{equation}
\begin{equation}
\hat{T}_{h-e} = \hat{V}_{h-e} + \hat{V}_{h-e} \hat{F}_3^D \hat{T}_{h-e},
\end{equation}
through the Faddeev equations \cite{Faddeev} 
\begin{displaymath}
\hat{S}_{h-h} = \hat{T}_{h-h} + \hat{T}_{h-h} \hat{F}_3^D \hat{S}_{h-e}, 
\end{displaymath}
\begin{equation}
\hat{S}_{h-e} = \hat{T}_{h-e} + \hat{T}_{h-e} \hat{F}_3^D \hat{S}_{h-h}.
\label{Faddeev}
\end{equation}
Inserting (\ref{Faddeev}) into  (\ref{f3}) one gets the expression  for the
3-particle resolvent in terms of scattering operators $S_{h-e}$ and
$T_{h-h}$ 
\begin{equation}
\label{ff3}
\hat{F}_3 = 
\hat{F}_3^D + 
\hat{F}_3^D  \left( \hat{T}_{h-h} + \hat{T}_{h-h}\hat{F}_3^D \hat{S}_{h-e} 
+ S_{h-e} \right)  \hat{F}_3^D.
\end{equation}
In this expression $\hat{F}_3^D$ and 
$\hat{T}$-operators -or rather their matrix 
elements between three-particle states- have an analytical expression, 
while the inclusion of scattering operator $\hat{S}_{h-e}$ will require the 
solution of an integral equation. 

After some algebra the hole propagator becomes
\begin{equation}
\label{ftt'}
{\cal G}^-\!\left({\bf k} n \sigma,\omega\right) =
- G_{ss}(z) = 
\frac{1}{\displaystyle \omega-E_0\left(N_e\right) +H^H_{ss} 
+ \sum_{tt'}F_{3tt'}V_{t's}V_{st}};
\end{equation}
with the notation $ G_{ss} \equiv \left<s\right|\hat{G}\left|s\right>$, 
$ F_{3 tt'}  \equiv 
\left<t\right|\hat{F}_3\left|t'\right> $ etc.. 
Since the difference between the ground state energy of the $N_e$-particle 
system $E_0(N_e)$ and the average of $\hat{H}^H$ 
over $\left|s\right>$ states turns out to be
\[
E_0(N_e) - H^H_{ss} = 
\epsilon_{{\bf k} n \sigma}^H + Q^n_{{\bf k} \sigma} = 
\epsilon_{{\bf k} n \sigma}^{MF},
\]
the mean field band eigenvalues appear naturally in the denominator of the hole
propagator. Comparing eq.~(\ref{ftt'}) with the usual expression
\begin{equation}   
{\cal G}^-\!\left({\bf k} n \sigma,\omega\right)=
\frac{1}{\omega-\epsilon_{{\bf k} n \sigma}^{MF}
-\Sigma^-_{{\bf k} n \sigma} (\omega) },
\label{selfenerg}
\end{equation}
we can identify the self-energy correction to band eigenvalues as 
\begin{equation}
\Sigma^-\!\left({\bf k} n \sigma,\omega\right) =
 - \sum_{tt'}F_{3tt'}V_{t's}V_{st}. 
\label{sigmab}
\end{equation}

The procedure we have outlined  ends up with a result which has a simple
physical interpretation: the creation of one hole in an unfilled valence
band is followed by multiple h-h and h-e scattering which is responsible
of a renormalization - through self-energy corrections - of the energy
states. The efficiency of the scattering processes depends a) on the
strength of the screened on-site e-e interaction and b) on the available
empty states (i.e. on the number of {\it initial} valence holes). This
explains the well known differences between various transition metals (Ni
and Cu, for instance) and possibly those arising in cuprates as a
consequence of hole doping. 

The self-energy $\Sigma^{+}$ and spectral function $D^{+}$ for electron 
addition can be calculated in the same way as described above just 
exchanging empty states with filled ones and vice versa,
as described in detail in ref.~\cite{Calandra}.

\subsection*{3BS at work}

In order to calculate the self-energy $\Sigma^-\left({\bf k} n
\sigma,\omega\right)$ according to (\ref{sigmab}) 
one has to perform summations over the 3-particle states 
$\left|t\right>$ involving ${\bf k}$-vector conservation; 
this is done within the so called {\it local} 
approximation \cite{Treglia80}
\[
\delta_{{\bf k}={\bf k}'}=
	\frac{1}{N}\sum_{\bf R}e^{i({\bf k}-{\bf k}')\cdot{\bf R}}
\simeq \frac{1}{N}.
\] 
This approximation 
allows to transform  the ${\bf k}$-vector summations into integrals 
involving the orbital density of states $n_{\alpha}\!(\epsilon)$
and to calculate T-matrices,
scattering operators, resolvents, self-energy and finally spectral 
functions according to the computational
strategy discussed in detail in ref. 
\cite{Calandra,Manghi97} which can be summarized as follows:
%
%
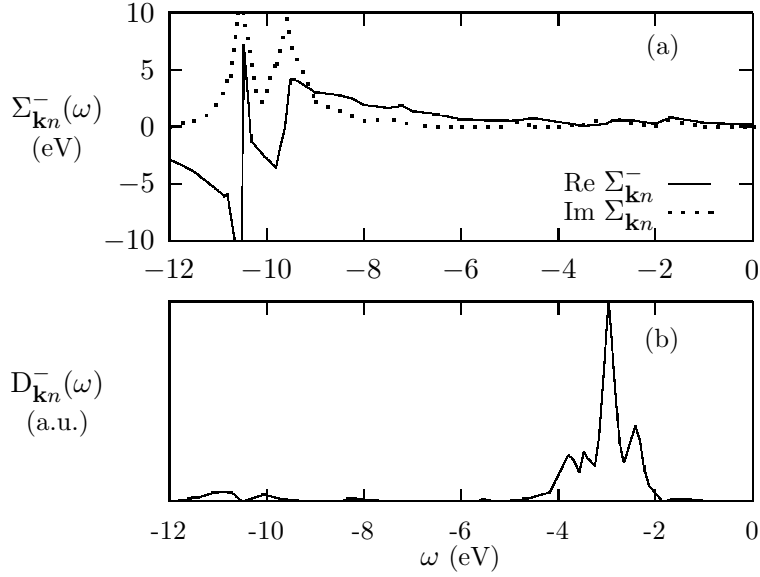
\begin{figure}[thb]
\begin{center}
\setlength{\unitlength}{0.240900pt}
\ifx\plotpoint\undefined\newsavebox{\plotpoint}\fi
\sbox{\plotpoint}{\rule[-0.200pt]{0.400pt}{0.400pt}}%
\begin{picture}(1200,450)(0,0)
\font\gnuplot=cmr10 at 10pt
\gnuplot
\sbox{\plotpoint}{\rule[-0.200pt]{0.400pt}{0.400pt}}%
\put(220.0,68.0){\rule[-0.200pt]{4.818pt}{0.400pt}}
\put(198,68){\makebox(0,0)[r]{$-10$}}
\put(1116.0,68.0){\rule[-0.200pt]{4.818pt}{0.400pt}}
\put(220.0,158.0){\rule[-0.200pt]{4.818pt}{0.400pt}}
\put(198,158){\makebox(0,0)[r]{$-5$}}
\put(1116.0,158.0){\rule[-0.200pt]{4.818pt}{0.400pt}}
\put(220.0,248.0){\rule[-0.200pt]{4.818pt}{0.400pt}}
\put(198,248){\makebox(0,0)[r]{$0$}}
\put(1116.0,248.0){\rule[-0.200pt]{4.818pt}{0.400pt}}
\put(220.0,337.0){\rule[-0.200pt]{4.818pt}{0.400pt}}
\put(198,337){\makebox(0,0)[r]{$5$}}
\put(1116.0,337.0){\rule[-0.200pt]{4.818pt}{0.400pt}}
\put(220.0,427.0){\rule[-0.200pt]{4.818pt}{0.400pt}}
\put(198,427){\makebox(0,0)[r]{$10$}}
\put(1116.0,427.0){\rule[-0.200pt]{4.818pt}{0.400pt}}
\put(220.0,68.0){\rule[-0.200pt]{0.400pt}{4.818pt}}
\put(220,23){\makebox(0,0){$-12$}}
\put(220.0,407.0){\rule[-0.200pt]{0.400pt}{4.818pt}}
\put(373.0,68.0){\rule[-0.200pt]{0.400pt}{4.818pt}}
\put(373,23){\makebox(0,0){$-10$}}
\put(373.0,407.0){\rule[-0.200pt]{0.400pt}{4.818pt}}
\put(525.0,68.0){\rule[-0.200pt]{0.400pt}{4.818pt}}
\put(525,23){\makebox(0,0){$-8$}}
\put(525.0,407.0){\rule[-0.200pt]{0.400pt}{4.818pt}}
\put(678.0,68.0){\rule[-0.200pt]{0.400pt}{4.818pt}}
\put(678,23){\makebox(0,0){$-6$}}
\put(678.0,407.0){\rule[-0.200pt]{0.400pt}{4.818pt}}
\put(831.0,68.0){\rule[-0.200pt]{0.400pt}{4.818pt}}
\put(831,23){\makebox(0,0){$-4$}}
\put(831.0,407.0){\rule[-0.200pt]{0.400pt}{4.818pt}}
\put(983.0,68.0){\rule[-0.200pt]{0.400pt}{4.818pt}}
\put(983,23){\makebox(0,0){$-2$}}
\put(983.0,407.0){\rule[-0.200pt]{0.400pt}{4.818pt}}
\put(1136.0,68.0){\rule[-0.200pt]{0.400pt}{4.818pt}}
\put(1136,23){\makebox(0,0){$0$}}
\put(1136.0,407.0){\rule[-0.200pt]{0.400pt}{4.818pt}}
\put(220.0,68.0){\rule[-0.200pt]{220.664pt}{0.400pt}}
\put(1136.0,68.0){\rule[-0.200pt]{0.400pt}{86.483pt}}
\put(220.0,427.0){\rule[-0.200pt]{220.664pt}{0.400pt}}
\put(45,247){\makebox(0,0){\shortstack{$\Sigma^-_{{\bf k}n}\!
\left(\omega\right)$\\(eV)}}}
\put(1022,373){\makebox(0,0)[r]{(a)}}
\put(220.0,68.0){\rule[-0.200pt]{0.400pt}{86.483pt}}
\put(983,158){\makebox(0,0)[r]{Re $\Sigma^-_{{\bf k}n}$}}
\put(1005.0,158.0){\rule[-0.200pt]{15.899pt}{0.400pt}}
\put(220,197){\usebox{\plotpoint}}
\multiput(220.00,195.92)(1.008,-0.495){35}{\rule{0.900pt}{0.119pt}}
\multiput(220.00,196.17)(36.132,-19.000){2}{\rule{0.450pt}{0.400pt}}
\multiput(258.00,176.92)(0.634,-0.497){57}{\rule{0.607pt}{0.120pt}}
\multiput(258.00,177.17)(36.741,-30.000){2}{\rule{0.303pt}{0.400pt}}
\multiput(296.00,146.92)(0.600,-0.491){17}{\rule{0.580pt}{0.118pt}}
\multiput(296.00,147.17)(10.796,-10.000){2}{\rule{0.290pt}{0.400pt}}
\multiput(308.60,138.00)(0.468,0.627){5}{\rule{0.113pt}{0.600pt}}
\multiput(307.17,138.00)(4.000,3.755){2}{\rule{0.400pt}{0.300pt}}
\multiput(312.58,132.21)(0.492,-3.210){21}{\rule{0.119pt}{2.600pt}}
\multiput(311.17,137.60)(12.000,-69.604){2}{\rule{0.400pt}{1.300pt}}
\multiput(335.61,68.00)(0.447,68.780){3}{\rule{0.108pt}{41.300pt}}
\multiput(334.17,68.00)(3.000,223.280){2}{\rule{0.400pt}{20.650pt}}
\multiput(338.58,355.55)(0.492,-6.528){21}{\rule{0.119pt}{5.167pt}}
\multiput(337.17,366.28)(12.000,-141.276){2}{\rule{0.400pt}{2.583pt}}
\multiput(350.58,222.64)(0.496,-0.587){43}{\rule{0.120pt}{0.570pt}}
\multiput(349.17,223.82)(23.000,-25.818){2}{\rule{0.400pt}{0.285pt}}
\multiput(373.00,196.92)(0.497,-0.494){27}{\rule{0.500pt}{0.119pt}}
\multiput(373.00,197.17)(13.962,-15.000){2}{\rule{0.250pt}{0.400pt}}
\multiput(388.58,183.00)(0.494,2.207){27}{\rule{0.119pt}{1.833pt}}
\multiput(387.17,183.00)(15.000,61.195){2}{\rule{0.400pt}{0.917pt}}
\multiput(403.59,248.00)(0.488,4.852){13}{\rule{0.117pt}{3.800pt}}
\multiput(402.17,248.00)(8.000,66.113){2}{\rule{0.400pt}{1.900pt}}
\multiput(418.00,320.92)(0.778,-0.496){37}{\rule{0.720pt}{0.119pt}}
\multiput(418.00,321.17)(29.506,-20.000){2}{\rule{0.360pt}{0.400pt}}
\multiput(449.00,300.93)(4.161,-0.477){7}{\rule{3.140pt}{0.115pt}}
\multiput(449.00,301.17)(31.483,-5.000){2}{\rule{1.570pt}{0.400pt}}
\multiput(487.00,295.93)(2.027,-0.482){9}{\rule{1.633pt}{0.116pt}}
\multiput(487.00,296.17)(19.610,-6.000){2}{\rule{0.817pt}{0.400pt}}
\multiput(510.00,289.93)(0.844,-0.489){15}{\rule{0.767pt}{0.118pt}}
\multiput(510.00,290.17)(13.409,-9.000){2}{\rule{0.383pt}{0.400pt}}
\multiput(525.00,280.93)(4.272,-0.477){7}{\rule{3.220pt}{0.115pt}}
\multiput(525.00,281.17)(32.317,-5.000){2}{\rule{1.610pt}{0.400pt}}
\multiput(564.00,277.60)(3.113,0.468){5}{\rule{2.300pt}{0.113pt}}
\multiput(564.00,276.17)(17.226,4.000){2}{\rule{1.150pt}{0.400pt}}
\multiput(586.00,279.93)(0.902,-0.489){15}{\rule{0.811pt}{0.118pt}}
\multiput(586.00,280.17)(14.316,-9.000){2}{\rule{0.406pt}{0.400pt}}
\multiput(602.00,270.93)(4.161,-0.477){7}{\rule{3.140pt}{0.115pt}}
\multiput(602.00,271.17)(31.483,-5.000){2}{\rule{1.570pt}{0.400pt}}
\multiput(640.00,265.93)(2.475,-0.488){13}{\rule{2.000pt}{0.117pt}}
\multiput(640.00,266.17)(33.849,-8.000){2}{\rule{1.000pt}{0.400pt}}
\put(678,257.67){\rule{9.154pt}{0.400pt}}
\multiput(678.00,258.17)(19.000,-1.000){2}{\rule{4.577pt}{0.400pt}}
\put(716,256.67){\rule{9.154pt}{0.400pt}}
\multiput(716.00,257.17)(19.000,-1.000){2}{\rule{4.577pt}{0.400pt}}
\multiput(754.00,257.60)(5.599,0.468){5}{\rule{4.000pt}{0.113pt}}
\multiput(754.00,256.17)(30.698,4.000){2}{\rule{2.000pt}{0.400pt}}
\multiput(793.00,259.93)(3.384,-0.482){9}{\rule{2.633pt}{0.116pt}}
\multiput(793.00,260.17)(32.534,-6.000){2}{\rule{1.317pt}{0.400pt}}
\multiput(831.00,253.93)(3.384,-0.482){9}{\rule{2.633pt}{0.116pt}}
\multiput(831.00,254.17)(32.534,-6.000){2}{\rule{1.317pt}{0.400pt}}
\multiput(869.00,249.61)(8.276,0.447){3}{\rule{5.167pt}{0.108pt}}
\multiput(869.00,248.17)(27.276,3.000){2}{\rule{2.583pt}{0.400pt}}
\multiput(907.00,252.59)(0.821,0.477){7}{\rule{0.740pt}{0.115pt}}
\multiput(907.00,251.17)(6.464,5.000){2}{\rule{0.370pt}{0.400pt}}
\put(915,256.67){\rule{7.227pt}{0.400pt}}
\multiput(915.00,256.17)(15.000,1.000){2}{\rule{3.613pt}{0.400pt}}
\multiput(945.00,256.93)(3.384,-0.482){9}{\rule{2.633pt}{0.116pt}}
\multiput(945.00,257.17)(32.534,-6.000){2}{\rule{1.317pt}{0.400pt}}
\multiput(983.00,252.58)(1.173,0.491){17}{\rule{1.020pt}{0.118pt}}
\multiput(983.00,251.17)(20.883,10.000){2}{\rule{0.510pt}{0.400pt}}
\put(1006,260.17){\rule{3.300pt}{0.400pt}}
\multiput(1006.00,261.17)(9.151,-2.000){2}{\rule{1.650pt}{0.400pt}}
\multiput(1022.00,258.93)(3.384,-0.482){9}{\rule{2.633pt}{0.116pt}}
\multiput(1022.00,259.17)(32.534,-6.000){2}{\rule{1.317pt}{0.400pt}}
\put(1060,252.17){\rule{7.700pt}{0.400pt}}
\multiput(1060.00,253.17)(22.018,-2.000){2}{\rule{3.850pt}{0.400pt}}
\put(411.0,322.0){\rule[-0.200pt]{1.686pt}{0.400pt}}
\put(1098.0,252.0){\rule[-0.200pt]{9.154pt}{0.400pt}}
\sbox{\plotpoint}{\rule[-0.500pt]{1.000pt}{1.000pt}}%
\put(983,113){\makebox(0,0)[r]{Im $\Sigma^-_{{\bf k}n}$}}
\multiput(1005,113)(20.756,0.000){4}{\usebox{\plotpoint}}
\put(1071,113){\usebox{\plotpoint}}
\put(220,248){\usebox{\plotpoint}}
\put(220.00,248.00){\usebox{\plotpoint}}
\put(240.11,253.11){\usebox{\plotpoint}}
\multiput(258,257)(19.034,8.276){2}{\usebox{\plotpoint}}
\put(292.33,282.10){\usebox{\plotpoint}}
\put(303.26,299.71){\usebox{\plotpoint}}
\put(309.67,319.42){\usebox{\plotpoint}}
\multiput(312,327)(20.756,0.000){0}{\usebox{\plotpoint}}
\multiput(315,327)(2.473,20.608){5}{\usebox{\plotpoint}}
\put(327,427){\usebox{\plotpoint}}
\multiput(335,427)(3.314,-20.489){7}{\usebox{\plotpoint}}
\put(363.38,287.01){\usebox{\plotpoint}}
\put(371.71,303.61){\usebox{\plotpoint}}
\put(376.89,323.68){\usebox{\plotpoint}}
\put(384.42,342.52){\usebox{\plotpoint}}
\multiput(388,347)(4.349,20.295){3}{\usebox{\plotpoint}}
\multiput(403,417)(3.633,-20.435){2}{\usebox{\plotpoint}}
\multiput(411,372)(8.471,-18.948){5}{\usebox{\plotpoint}}
\multiput(449,287)(19.306,-7.621){2}{\usebox{\plotpoint}}
\multiput(487,272)(19.306,-7.621){2}{\usebox{\plotpoint}}
\multiput(525,257)(20.749,0.532){2}{\usebox{\plotpoint}}
\multiput(564,258)(20.756,0.000){0}{\usebox{\plotpoint}}
\multiput(571,258)(20.097,-5.186){2}{\usebox{\plotpoint}}
\put(620.46,249.03){\usebox{\plotpoint}}
\multiput(640,248)(20.756,0.000){2}{\usebox{\plotpoint}}
\multiput(678,248)(20.756,0.000){4}{\usebox{\plotpoint}}
\put(763.98,254.14){\usebox{\plotpoint}}
\put(783.45,250.94){\usebox{\plotpoint}}
\multiput(793,248)(20.756,0.000){2}{\usebox{\plotpoint}}
\multiput(831,248)(20.691,1.634){2}{\usebox{\plotpoint}}
\multiput(869,251)(19.937,5.771){2}{\usebox{\plotpoint}}
\multiput(907,262)(19.690,-6.563){0}{\usebox{\plotpoint}}
\put(925.41,258.70){\usebox{\plotpoint}}
\put(943.39,252.96){\usebox{\plotpoint}}
\put(963.82,250.51){\usebox{\plotpoint}}
\multiput(983,249)(18.564,9.282){2}{\usebox{\plotpoint}}
\multiput(1003,259)(20.072,-5.282){0}{\usebox{\plotpoint}}
\put(1023.00,253.80){\usebox{\plotpoint}}
\put(1043.43,250.16){\usebox{\plotpoint}}
\multiput(1060,248)(20.756,0.000){4}{\usebox{\plotpoint}}
\put(1136,248){\usebox{\plotpoint}}
\end{picture}
\setlength{\unitlength}{0.240900pt}
\ifx\plotpoint\undefined\newsavebox{\plotpoint}\fi
\sbox{\plotpoint}{\rule[-0.200pt]{0.400pt}{0.400pt}}%
\begin{picture}(1200,450)(0,0)
\font\gnuplot=cmr10 at 10pt
\gnuplot
\sbox{\plotpoint}{\rule[-0.200pt]{0.400pt}{0.400pt}}%
\put(220.0,113.0){\rule[-0.200pt]{0.400pt}{4.818pt}}
\put(220,68){\makebox(0,0){-12}}
\put(220.0,407.0){\rule[-0.200pt]{0.400pt}{4.818pt}}
\put(373.0,113.0){\rule[-0.200pt]{0.400pt}{4.818pt}}
\put(373,68){\makebox(0,0){-10}}
\put(373.0,407.0){\rule[-0.200pt]{0.400pt}{4.818pt}}
\put(525.0,113.0){\rule[-0.200pt]{0.400pt}{4.818pt}}
\put(525,68){\makebox(0,0){-8}}
\put(525.0,407.0){\rule[-0.200pt]{0.400pt}{4.818pt}}
\put(678.0,113.0){\rule[-0.200pt]{0.400pt}{4.818pt}}
\put(678,68){\makebox(0,0){-6}}
\put(678.0,407.0){\rule[-0.200pt]{0.400pt}{4.818pt}}
\put(831.0,113.0){\rule[-0.200pt]{0.400pt}{4.818pt}}
\put(831,68){\makebox(0,0){-4}}
\put(831.0,407.0){\rule[-0.200pt]{0.400pt}{4.818pt}}
\put(983.0,113.0){\rule[-0.200pt]{0.400pt}{4.818pt}}
\put(983,68){\makebox(0,0){-2}}
\put(983.0,407.0){\rule[-0.200pt]{0.400pt}{4.818pt}}
\put(1136.0,113.0){\rule[-0.200pt]{0.400pt}{4.818pt}}
\put(1136,68){\makebox(0,0){0}}
\put(1136.0,407.0){\rule[-0.200pt]{0.400pt}{4.818pt}}
\put(220.0,113.0){\rule[-0.200pt]{220.664pt}{0.400pt}}
\put(1136.0,113.0){\rule[-0.200pt]{0.400pt}{75.643pt}}
\put(220.0,427.0){\rule[-0.200pt]{220.664pt}{0.400pt}}
\put(45,270){\makebox(0,0){\shortstack{${\rm D}^-_{{\bf k}n}\!
\left(\omega\right)$\\(a.u.)}}}
\put(678,23){\makebox(0,0){$\omega$ (eV)}}
\put(1022,366){\makebox(0,0)[r]{(b)}}
\put(220.0,113.0){\rule[-0.200pt]{0.400pt}{75.643pt}}
\put(220,113){\usebox{\plotpoint}}
\multiput(220.00,113.59)(3.827,0.477){7}{\rule{2.900pt}{0.115pt}}
\multiput(220.00,112.17)(28.981,5.000){2}{\rule{1.450pt}{0.400pt}}
\put(255,118.17){\rule{2.900pt}{0.400pt}}
\multiput(255.00,117.17)(7.981,2.000){2}{\rule{1.450pt}{0.400pt}}
\multiput(269.00,120.59)(1.560,0.485){11}{\rule{1.300pt}{0.117pt}}
\multiput(269.00,119.17)(18.302,7.000){2}{\rule{0.650pt}{0.400pt}}
\put(290,126.67){\rule{3.614pt}{0.400pt}}
\multiput(290.00,126.17)(7.500,1.000){2}{\rule{1.807pt}{0.400pt}}
\put(305,126.67){\rule{3.373pt}{0.400pt}}
\multiput(305.00,127.17)(7.000,-1.000){2}{\rule{1.686pt}{0.400pt}}
\multiput(319.00,125.92)(0.570,-0.494){25}{\rule{0.557pt}{0.119pt}}
\multiput(319.00,126.17)(14.844,-14.000){2}{\rule{0.279pt}{0.400pt}}
\multiput(335.00,113.58)(1.534,0.492){19}{\rule{1.300pt}{0.118pt}}
\multiput(335.00,112.17)(30.302,11.000){2}{\rule{0.650pt}{0.400pt}}
\multiput(368.00,122.93)(2.094,-0.485){11}{\rule{1.700pt}{0.117pt}}
\multiput(368.00,123.17)(24.472,-7.000){2}{\rule{0.850pt}{0.400pt}}
\multiput(396.00,115.95)(7.607,-0.447){3}{\rule{4.767pt}{0.108pt}}
\multiput(396.00,116.17)(25.107,-3.000){2}{\rule{2.383pt}{0.400pt}}
\put(431,112.67){\rule{8.672pt}{0.400pt}}
\multiput(431.00,113.17)(18.000,-1.000){2}{\rule{4.336pt}{0.400pt}}
\multiput(488.00,113.59)(1.489,0.477){7}{\rule{1.220pt}{0.115pt}}
\multiput(488.00,112.17)(11.468,5.000){2}{\rule{0.610pt}{0.400pt}}
\multiput(502.00,116.93)(7.723,-0.477){7}{\rule{5.700pt}{0.115pt}}
\multiput(502.00,117.17)(58.169,-5.000){2}{\rule{2.850pt}{0.400pt}}
\put(467.0,113.0){\rule[-0.200pt]{5.059pt}{0.400pt}}
\put(643,112.67){\rule{14.936pt}{0.400pt}}
\multiput(643.00,112.17)(31.000,1.000){2}{\rule{7.468pt}{0.400pt}}
\multiput(705.00,114.61)(1.579,0.447){3}{\rule{1.167pt}{0.108pt}}
\multiput(705.00,113.17)(5.579,3.000){2}{\rule{0.583pt}{0.400pt}}
\multiput(713.00,115.94)(2.967,-0.468){5}{\rule{2.200pt}{0.113pt}}
\multiput(713.00,116.17)(16.434,-4.000){2}{\rule{1.100pt}{0.400pt}}
\multiput(734.00,113.59)(5.497,0.477){7}{\rule{4.100pt}{0.115pt}}
\multiput(734.00,112.17)(41.490,5.000){2}{\rule{2.050pt}{0.400pt}}
\multiput(784.00,118.58)(1.798,0.491){17}{\rule{1.500pt}{0.118pt}}
\multiput(784.00,117.17)(31.887,10.000){2}{\rule{0.750pt}{0.400pt}}
\multiput(819.58,128.00)(0.497,1.023){53}{\rule{0.120pt}{0.914pt}}
\multiput(818.17,128.00)(28.000,55.102){2}{\rule{0.400pt}{0.457pt}}
\multiput(847.00,183.94)(0.920,-0.468){5}{\rule{0.800pt}{0.113pt}}
\multiput(847.00,184.17)(5.340,-4.000){2}{\rule{0.400pt}{0.400pt}}
\multiput(854.58,177.26)(0.492,-1.015){19}{\rule{0.118pt}{0.900pt}}
\multiput(853.17,179.13)(11.000,-20.132){2}{\rule{0.400pt}{0.450pt}}
\multiput(865.59,159.00)(0.485,2.476){11}{\rule{0.117pt}{1.986pt}}
\multiput(864.17,159.00)(7.000,28.879){2}{\rule{0.400pt}{0.993pt}}
\multiput(872.59,188.50)(0.485,-0.950){11}{\rule{0.117pt}{0.843pt}}
\multiput(871.17,190.25)(7.000,-11.251){2}{\rule{0.400pt}{0.421pt}}
\multiput(879.00,177.92)(0.495,-0.491){17}{\rule{0.500pt}{0.118pt}}
\multiput(879.00,178.17)(8.962,-10.000){2}{\rule{0.250pt}{0.400pt}}
\multiput(889.59,169.00)(0.485,3.391){11}{\rule{0.117pt}{2.671pt}}
\multiput(888.17,169.00)(7.000,39.455){2}{\rule{0.400pt}{1.336pt}}
\multiput(896.58,214.00)(0.494,7.269){27}{\rule{0.119pt}{5.780pt}}
\multiput(895.17,214.00)(15.000,201.003){2}{\rule{0.400pt}{2.890pt}}
\multiput(911.58,404.41)(0.494,-6.850){25}{\rule{0.119pt}{5.443pt}}
\multiput(910.17,415.70)(14.000,-175.703){2}{\rule{0.400pt}{2.721pt}}
\multiput(925.61,219.66)(0.447,-7.830){3}{\rule{0.108pt}{4.900pt}}
\multiput(924.17,229.83)(3.000,-25.830){2}{\rule{0.400pt}{2.450pt}}
\multiput(928.59,196.47)(0.485,-2.247){11}{\rule{0.117pt}{1.814pt}}
\multiput(927.17,200.23)(7.000,-26.234){2}{\rule{0.400pt}{0.907pt}}
\multiput(935.58,174.00)(0.495,1.632){33}{\rule{0.119pt}{1.389pt}}
\multiput(934.17,174.00)(18.000,55.117){2}{\rule{0.400pt}{0.694pt}}
\multiput(953.59,225.89)(0.485,-1.789){11}{\rule{0.117pt}{1.471pt}}
\multiput(952.17,228.95)(7.000,-20.946){2}{\rule{0.400pt}{0.736pt}}
\multiput(960.59,197.15)(0.485,-3.315){11}{\rule{0.117pt}{2.614pt}}
\multiput(959.17,202.57)(7.000,-38.574){2}{\rule{0.400pt}{1.307pt}}
\multiput(967.59,157.42)(0.485,-1.942){11}{\rule{0.117pt}{1.586pt}}
\multiput(966.17,160.71)(7.000,-22.709){2}{\rule{0.400pt}{0.793pt}}
\multiput(974.58,135.69)(0.496,-0.571){39}{\rule{0.119pt}{0.557pt}}
\multiput(973.17,136.84)(21.000,-22.844){2}{\rule{0.400pt}{0.279pt}}
\multiput(995.00,114.61)(3.811,0.447){3}{\rule{2.500pt}{0.108pt}}
\multiput(995.00,113.17)(12.811,3.000){2}{\rule{1.250pt}{0.400pt}}
\put(1013,115.67){\rule{4.095pt}{0.400pt}}
\multiput(1013.00,116.17)(8.500,-1.000){2}{\rule{2.048pt}{0.400pt}}
\put(1030,114.67){\rule{5.059pt}{0.400pt}}
\multiput(1030.00,115.17)(10.500,-1.000){2}{\rule{2.529pt}{0.400pt}}
\put(1051,113.17){\rule{3.100pt}{0.400pt}}
\multiput(1051.00,114.17)(8.566,-2.000){2}{\rule{1.550pt}{0.400pt}}
\put(572.0,113.0){\rule[-0.200pt]{17.104pt}{0.400pt}}
\put(1066.0,113.0){\rule[-0.200pt]{16.863pt}{0.400pt}}
\end{picture}
\end{center}
\caption{Real (solid line) and imaginary (dotted line) part of self-energy
$\Sigma^{-}_{{\bf k}n}(\omega)$ (a) and spectral function (b) 
in paramagnetic CuGeO$_3$ for the
creation of a hole at the $\Gamma$ point corresponding to the
single particle band energy $\epsilon^{n}_{\bf k}=$~-3.58~eV. Energies are
referred to E$_f$. 
\label{fig_2}}
\end{figure}
\begin{itemize}
\item {\tt Input:
  band structure ($\epsilon_k^n, C^n_{ \alpha \uparrow}(k),
n_{\alpha \uparrow}(\epsilon)$) and  $U$;}
\item {\tt free propagators}
\[
g_{h-h}^{\alpha  \beta}(\omega) =
\int_{-\infty}^{E_f}\!\!{\rm d\,} \epsilon' \int_{-\infty}^{E_f}
\!\! {\rm d\,} \epsilon
 \frac{{n}_{\alpha \downarrow}(\epsilon) {n}_{\beta \uparrow}(\epsilon')}
{\omega -\epsilon' -\epsilon - i\delta},
\]
\[
g_{h-e}^{\alpha\beta}(\omega)=\int_{-\infty}^{E_f}\!\!{\rm d\,}\epsilon'
\int_{E_f}^{\infty}\!\!{\rm d\,}\epsilon
 \frac{{n}_{\alpha \downarrow}(\epsilon) {n}_{ \beta \uparrow}(\epsilon')}
{\omega -\epsilon' +\epsilon - i\delta},
\qquad
g^{\beta}\!(\omega) = \int_{-\infty}^{E_f}\!\!{\rm d\,}\epsilon'
 \frac{{n}_{\beta \uparrow}(\epsilon')} {\omega -\epsilon' - i\delta};
\]
\item {\tt T-matrices}
\[
T_{h-h}^{\alpha \beta}(\omega) = \frac {{U}} {1 + U
 g_{h-h}^{\alpha \beta}(\omega)},\qquad
T_{h-e}^{\alpha \beta}(\omega) = \frac {-{U}} {1 - U
 g_{h-e}^{\alpha  \beta}(\omega)};
\]
\item {\tt kernel}
\[
K^{\alpha \beta}\!(\omega,\epsilon,\epsilon') =
\int_{-\infty}^{E_f}\!\!{\rm d\,}\epsilon''n_{\alpha \downarrow}(\epsilon''
)
g^{ \beta}\!(\omega+\epsilon''-\epsilon)
g^{ \beta}\!(\omega+\epsilon''-\epsilon')
T_{h-e}^{\alpha \beta}(\omega+\epsilon'')
T_{h-h}^{\alpha \beta}(\omega-\epsilon''),
\]
{\tt and}
\begin{eqnarray*}
B^{\alpha \beta}\!(\omega,\epsilon) &=&
\int_{-\infty}^{E_f}\!\!{\rm d\,}\epsilon'n_{\alpha \downarrow}(\epsilon')
g^{ \beta}\!(\omega+\epsilon'-\epsilon)
T_{h-e}^{\alpha \beta}(\omega+\epsilon')\\
&\times&
\left[g_{h-e}^{\alpha \beta}(\omega-\epsilon') +
\int_{E_f}^{\infty}\!\!{\rm d\,}
\epsilon''   n_{\alpha \downarrow} (\epsilon'')
 g^{ \beta}\!(\omega+\epsilon'-\epsilon'') g_{h-h}^{\alpha
\beta}(\omega-\epsilon'')
T_{h-h}^{\alpha \beta}(\omega-\epsilon'') \right];
\end{eqnarray*}
\item {\tt solve the integral equation}
\[
A^{\alpha \beta}\!(\omega,\epsilon) =
B^{\alpha \beta}\!(\omega,\epsilon) +
\int_{E_f}^{\infty}\!\!{\rm d\,} \epsilon'
n_{\alpha \downarrow} (\epsilon')
K^{\alpha \beta}\!(\omega,\epsilon,\epsilon')
A^{\alpha \beta}\!(\omega,\epsilon');
\]
\item {\tt orbital self-energy}
\[
\Sigma^-_{ \beta \uparrow}(\omega) = \sum_{\alpha}
\int_{E_f}^{\infty}\!\! {\rm d\,} \epsilon\,
n_{\alpha \downarrow}(\epsilon)  T_{h-h}^{\alpha  \beta} (\omega-\epsilon)
\left[1+ U A^{\alpha  \beta}\!(\omega \epsilon)\right];
\]
\item {\tt ${\bf K}$- and band-index dependent self-energy}
\[
{\Sigma}^-\!\left(k n \uparrow,\omega\right) =
U  \sum_{ \beta} |C^n_{ \beta \uparrow}( k)|^2
\left[\sum_{\alpha}
\int_{-\infty}^{E_f}\!\!{\rm d\,}
\epsilon\,  n_{\alpha \downarrow}(\epsilon)
\Sigma^-_{ \beta \uparrow}(\omega)\right];
\]
\item {\tt spectral function}
\[
D^{(-)}_{\uparrow}\!(\omega) =  \frac{1}{\pi} \sum_{ k n}{\rm Im}
 \frac{1} {\omega - \epsilon_k^n
-{\Sigma}^-(k n\uparrow, \omega)}.
\]
\end{itemize}

%
%
\begin{figure}[thb]
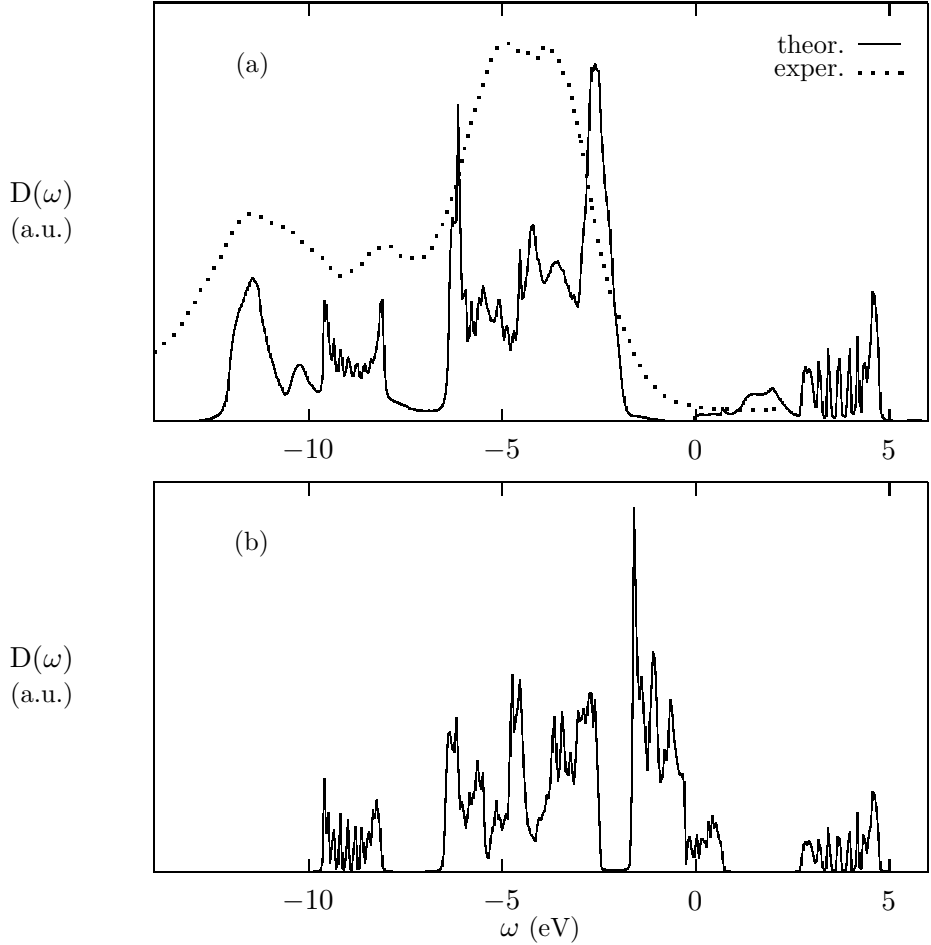

\begin{center}
\setlength{\unitlength}{0.240900pt}
\ifx\plotpoint\undefined\newsavebox{\plotpoint}\fi
\sbox{\plotpoint}{\rule[-0.200pt]{0.400pt}{0.400pt}}%

\end{center}
\caption{
Total spectral function \protect{${\rm D}\!\left(\omega\right)=
{\rm D}^-\!\left(\omega\right)+{\rm D}^+\!\left(\omega\right)$} 
for CuGeO$_3$ calculated  
by (a) 3BS theory with U=8~eV; (b) single particle density of states 
(Mattheiss, PRB {\bf 49,} 14050 (1994)). 
E$_f$ corresponds to $\omega=0$~eV.
}\label{fig_3}
\end{figure}
We report the results for a transition metal (nickel) and a cuprate 
(CuGeO$_3$) obtained by considering the interaction between opposite spin 
electrons localized on the transition metal sites as the dominant contribution, 
i.e.  
\begin{displaymath}
U_{\alpha \beta} = 
\cases{U_{d d} & for $\alpha,\beta=d$ orbitals \cr
0 & elsewhere; \cr}
\end{displaymath}
\begin{displaymath}
U_{\alpha\beta}-J_{\alpha\beta} \simeq 0.
\end{displaymath}
To apply this method to CuGeO$_3$ we have used the 
eigenstates/eigenvalues of ref. \cite{Mattheiss} and assumed $U_{d d}=8$~eV.
Fig.~\ref{fig_2} shows the hole spectral function  and self-energy for a
particular 
eigenstate ($\epsilon_{\bf k}^{n}$=-3.58~eV at the $\Gamma$ point). 
The peaks in the spectral function can be
classified either as quasiparticle excitations or as satellites, according
to the value of the imaginary part of the self-energy in the region of the
peak: quasi-particle excitations correspond to small imaginary part, and
give rise to the coherent part of the spectral function. Satellites occur
where the imaginary part of self-energy is large and correspond to 
short-lived excitations with a large intrinsic line-width;
we refer to them as to the incoherent part of the spectral 
function. 

%
%
\begin{figure}[thb]
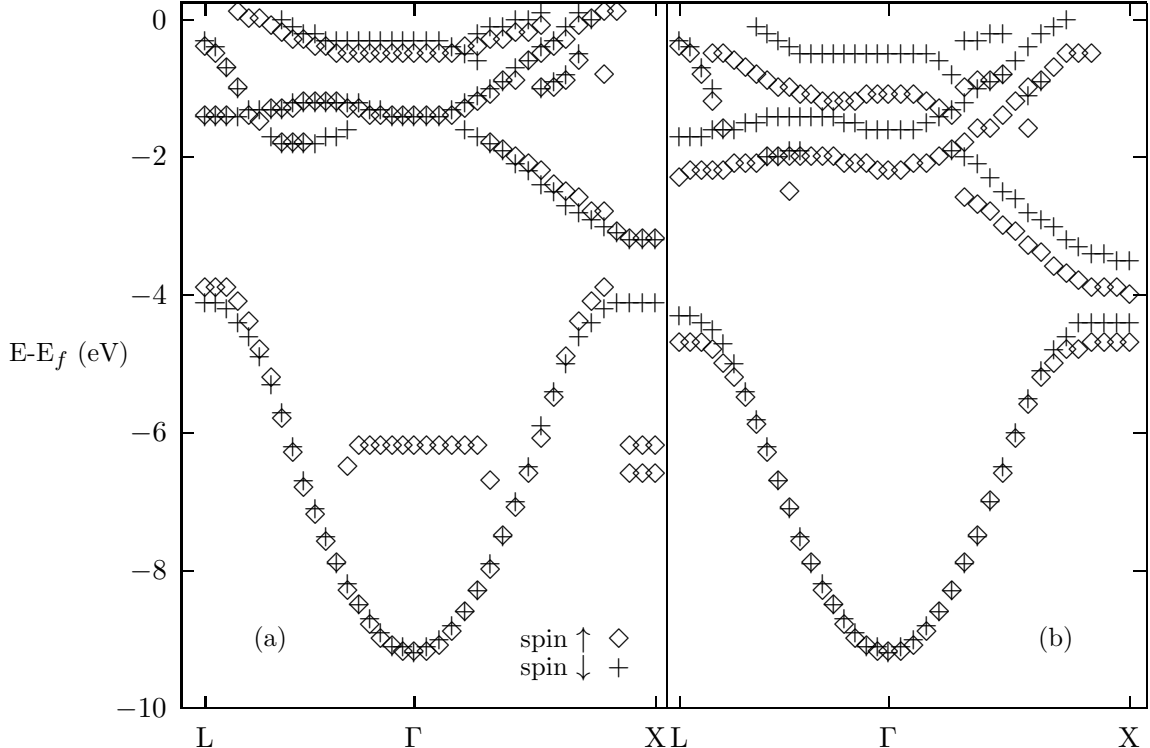

\begin{center}
\setlength{\unitlength}{0.240900pt}
\ifx\plotpoint\undefined\newsavebox{\plotpoint}\fi
\sbox{\plotpoint}{\rule[-0.200pt]{0.400pt}{0.400pt}}%

\end{center}
\caption{(a) Energy position of the peaks of nickel
in the spectral function at ${\bf k}$-points 
along high symmetry lines of the Brillouin Zone.  
(b) Single particle band structure 
(Manghi {\em et al.,} PRB {\bf 56,} 7149 (1997)).
}\label{fig_4}
\end{figure}
The hole and electron spectral functions 
are plotted in fig.~\ref{fig_3} and compared with single particle results. 
The effect
of electron correlation  on single-particle states
is dramatic: some bands are shifted to higher
binding energies, spectral weight is removed from the upper part of
the spectrum, and many new states (satellites) appear; 
only states around $-8\div -9$~eV and $-5\div -6$~eV are
practically unaffected being mainly  Ge and O derived. 
CuGeO$_3$ is an insulator but it is predicted to be a metal by
single particle band calculation; the inclusion of electron correlation
reproduces this insulating behavior and the energy gap, calculated
as the energy separation between electron removal and electron addition
spectra, reproduces the experimental one \cite{Villafiorita}. The
same was proven to be true also in the case of NiO, where a 3BS description
of Hubbard correlation was able to reproduced both the complex satellite
structure and the measured value of the insulating gap \cite{Manghi94}. 

The ability of 3BS approach to open up Hubbard gaps, i.e.~to reproduce an
insulating behaviour in a system which is metallic according to band
theory, is related to its non-perturbative character; the method can be
applied in any correlation regime and it has been shown
\cite{Igarashi,Manghi94,RontaniTh} to reproduce for U much
larger than the band width W ($\frac{U}{W} \rightarrow \infty$) the so
called ``Hubbard I'' solution \cite{HubbardI} of Hubbard Hamiltonian,
i.e.~the {\em atomic limit} where  hole and electron states are separated
by a Mott-Hubbard gap equal to U \cite{VillaTh}. 

We describe now the results of the application of 3BS to nickel. In this
case the on-site e-e repulsion is more 
effectively screened and the estimated
value of $U_{d d}$ is $\simeq$ 2 eV \cite{Springer97}. 
Fig.~\ref{fig_4}, reporting
the comparison between quasi-particle states and single particle ones,
shows that e-e correlation effects are still sizable and they are actually
essential in order to to reproduce the observed spectroscopical features,
i.e.~satellite structure at 6 eV binding energy, correct band width
(overestimated in LDA), exchange splitting \cite{Kreutz}, and energy dispersion
\cite{Manghi97}.

\subsection*{Summary and outlook}

We have described a method to include on-site interactions in the
description of hole and electrons states: ab-initio single-particle band
states are used as input mean-field eigenstates for the calculation of
self-energy corrections according to a 3-body scattering (3BS) solution of
a multi-orbital Hubbard Hamiltonian. When applied to valence states of
ferromagnetic nickel it allows to get a quasiparticle band structure which
compares much more favorably with the experimental observation than
conventional mean-field LDA, reproducing the observed band width, the
energy dispersion, the satellite structure and the exchange splitting.
Since the method does not rely on a perturbation expansion it has a wide
range of application, including any correlation regime. In the case of a
highly correlated system such as CuGeO$_3$, 3BS is able to reproduce both
the insulating behaviour and a correct overall picture of photoemission
experiments. 

Our present choice of empirically determining the parameter U of the
Hubbard Hamiltonian - which has been fixed to reproduce the satellite
binding energy - ensures that we obtain a good agreement with 
experiments; however  the possibility of reproducing both the satellite
structures and other spectroscopical features such as energy dispersion and
spin dependence in nickel and the insulating gap in CuGeO$_3$ can be seen
as a non trivial result and a success of the method itself: previous
methods based on a simplified description of the scattering channels
\cite{Liebsch,Manghi-ni} in fact have not been able to reproduce at the
same time the satellite energy position and the valence band width which
turned out to be systematically overestimated for values of the Coulomb
integral fixed to reproduce the satellite binding energy. The problem of
extracting Hubbard U from ab-initio calculation, either in the so called
Constrained-Density Functional scheme \cite{Dederichs} or as screened
Coulomb interaction \cite{Springer97,QD}, is an important issue which goes
in the direction of a full match between model Hamiltonians and realistic
systems and that we are presently 
considering as an implementation of our approach.

\end{document}